\documentclass[twocolumn,aps,prb,longbibliography,notitlepage]{revtex4-2}
\pdfoutput=1
\usepackage{graphicx}
\usepackage{float}
\usepackage{color}
\usepackage{amsmath} 
\usepackage{amssymb}
\usepackage{amsfonts}
\usepackage{upgreek}
\usepackage[normalem]{ulem}
\usepackage{comment}
\usepackage[euler]{textgreek}
\usepackage{lineno}
\usepackage[FIGTOPCAP]{subfigure}
\usepackage[bookmarks=false]{hyperref}
\usepackage[dvipsnames]{xcolor}
\hypersetup{pdfborder={0 0 0},
    colorlinks=true,
    linkcolor=blue,
    citecolor=blue,
    urlcolor=blue}
\newcommand{\stkout}[1]{\ifmmode\text{\sout{\ensuremath{#1}}}\else\sout{#1}\fi}
\newcommand{\bs}{\boldsymbol}
\newcommand{\pd}{\partial}
\begin{document}

\title{Spin anomalous-Hall unidirectional magnetoresistance}
\author{M. Mehraeen}
\affiliation{Department of Physics, Case Western Reserve University, Cleveland, OH 44106, USA}
\author{Steven S.-L. Zhang}
\thanks{E-mail: shulei.zhang@case.edu}
\affiliation{Department of Physics, Case Western Reserve University, Cleveland, OH 44106, USA}

\date{\today}

\begin{abstract}

We predict a spin anomalous-Hall unidirectional magnetoresistance (AH-UMR) in conducting bilayers composed of a ferromagnetic layer and a nonmagnetic layer, which does \textit{not} rely on the spin Hall effect in the normal metal layer|in stark contrast to the well-studied unidirectional spin-Hall magnetoresistance|but, instead, arises from the spin anomalous Hall effect in the ferromagnetic layer. Physically, it is the charge-spin conversion induced by the spin anomalous Hall effect that conspires with the structural inversion asymmetry to generate a net nonequilibrium spin density in the ferromagnetic layer, which, in turn, modulates the resistance of the bilayer when the direction of the applied current or the magnetization is reversed. The dependences of the spin AH-UMR effect on materials and geometric parameters are analyzed and compared with other nonlinear magnetoresistances. In particular, we show that, in magnetic bilayers where spin anomalous Hall and spin Hall effects are comparable, the overall UMR may undergo a sign change when the thickness of either layer is varied, suggesting a scheme to quantify the spin Hall or spin anomalous Hall angle via a nonlinear transport measurement. 

\end{abstract}

\maketitle

\section{Introduction}

Originating from the interplay between magnetism and relativistic spin-orbit interaction, the anomalous Hall (AH) effect in solid-state systems with broken time-reversal symmetry has been of enduring interest for more than a century~\cite{Hall1881AHE,Nagaosa10RMP}. One class of materials that has received particular attention in studies of this effect are conducting ferromagnets~\cite{Nagaosa10RMP}, such as ferromagnetic metals. 

The AH effect in ferromagnetic metals has several salient properties. Due to the coupling of spin and orbital degrees of freedom, the effect not only generates a transverse charge current|which is perpendicular to both the magnetization and the applied electric field, but also gives rise to a transverse spin current. And, in ferromagnetic metals with strong exchange interaction, conduction-electron spins are well aligned with the local magnetization, making the coupled spin and charge currents 
controllable by varying the direction of the magnetization. Furthermore, the mobility of conduction electrons in a ferromagnetic metal is, in general, spin-dependent, enabling mutual conversion between spin and charge currents mediated by the AH effect. 

These properties have been shown to spawn unconventional magnetoresistances in the linear response regime. For instance, both a bulk anisotropic magnetoresistance and planar Hall resistance may result from two consecutive transverse scatterings of spin-polarized conduction electrons, due to the AH effect~\cite{sZhang14JAP_AMR}. In geometrically confined systems|such as ferromagnetic-metal thin films or layered structures, the anomalous-Hall induced anisotropic magnetoresistances may acquire distinctive angular dependences, owing to the modulation of the bulk spin and charge currents caused by interfacial spin accumulation and the resulting diffusive spin current~\cite{zhang14Thesis,Tomo15PRAppl_AHE-SOT, xrWang16EPL_AMR,Tomo16PRB_AH-AMR,yhWu18NC_AHMR,yzWu20NJP_AHMR}. 

In the nonlinear response regime|where the Onsager reciprocal relations no longer hold, the role of the AH effect is yet to be explored. In this work, we unveil a unidirectional magnetoresistance (UMR) driven by the spin AH effect in conducting ferromagnet$\mid$nonmagnet bilayer systems, whereby the resistance can be altered by reversing the direction of either the magnetization or the applied electric field. Hereafter, we shall refer to this nonlinear magnetoresistance as spin anomalous-Hall unidirectional magnetoresistance (AH-UMR). 

The underlying physics of the spin AH-UMR can be understood as follows. In a single ferromagnetic-metal layer, the spin current induced by the spin AH effect creates spin accumulations of opposite orientations at the top and bottom surfaces, but there is no net nonequilibrium spin density due to inversion symmetry, as shown schematically in Fig.\,\ref{fig1}\textcolor{blue}{a}. This, however, is no longer the case when a nonmagnetic-metal layer is attached to the ferromagnetic-metal layer, as the spin accumulation at the interface would ``leak" into the nonmagnetic layer, resulting in a net nonequilibrium spin density in the ferromagnetic layer, which conspires with the spin asymmetry in electron mobility to produce the spin AH-UMR effect, as depicted in Figs.\,\ref{fig1}\textcolor{blue}{b} and\,\ref{fig1}\textcolor{blue}{c}.  

There is a key difference between the spin AH-UMR and other types of UMR effects previously studied in various magnetic systems~\cite{avci2015natphys,Jungwirth15PRB_UMR-DMS, shulei2016prb,Ferguson16APL_UMR-STO,yasuda2016prl, avci2018prl,lv2018natcomm,pnHai19JAP_UMR_TI-DMS, guillet2021prb,dWu21PRL_magnon-USMR}: for the spin AH-UMR, the nonequilibrium spin density in the presence of an external electric field emanates from the spin AH effect in the ferromagnetic layer itself, whereas for other UMRs|such as the unidirectional spin Hall magnetoresistance (USMR)~\cite{avci2015natphys} and the unidirectional Rashba magnetoresistance~\cite{guillet2021prb}|the nonequilibrium spin density is engendered 
by the spin Hall (SH) effect in the nonmagnetic layer~\cite{avci2015natphys,Jungwirth15PRB_UMR-DMS, shulei2016prb,Ferguson16APL_UMR-STO,avci2018prl,dWu21PRL_magnon-USMR} or by the Rashba-Edelstein effect due to the spin-momentum locked surface states~\cite{yasuda2016prl,lv2018natcomm,pnHai19JAP_UMR_TI-DMS,guillet2021prb,mandela21_QUMR}. 
\begin{figure*}[htp]
\label{fig1}
\centering 
\vspace*{-.4cm}
\hspace{0.3cm}
\subfigure[]{
\includegraphics[width=.3\linewidth]{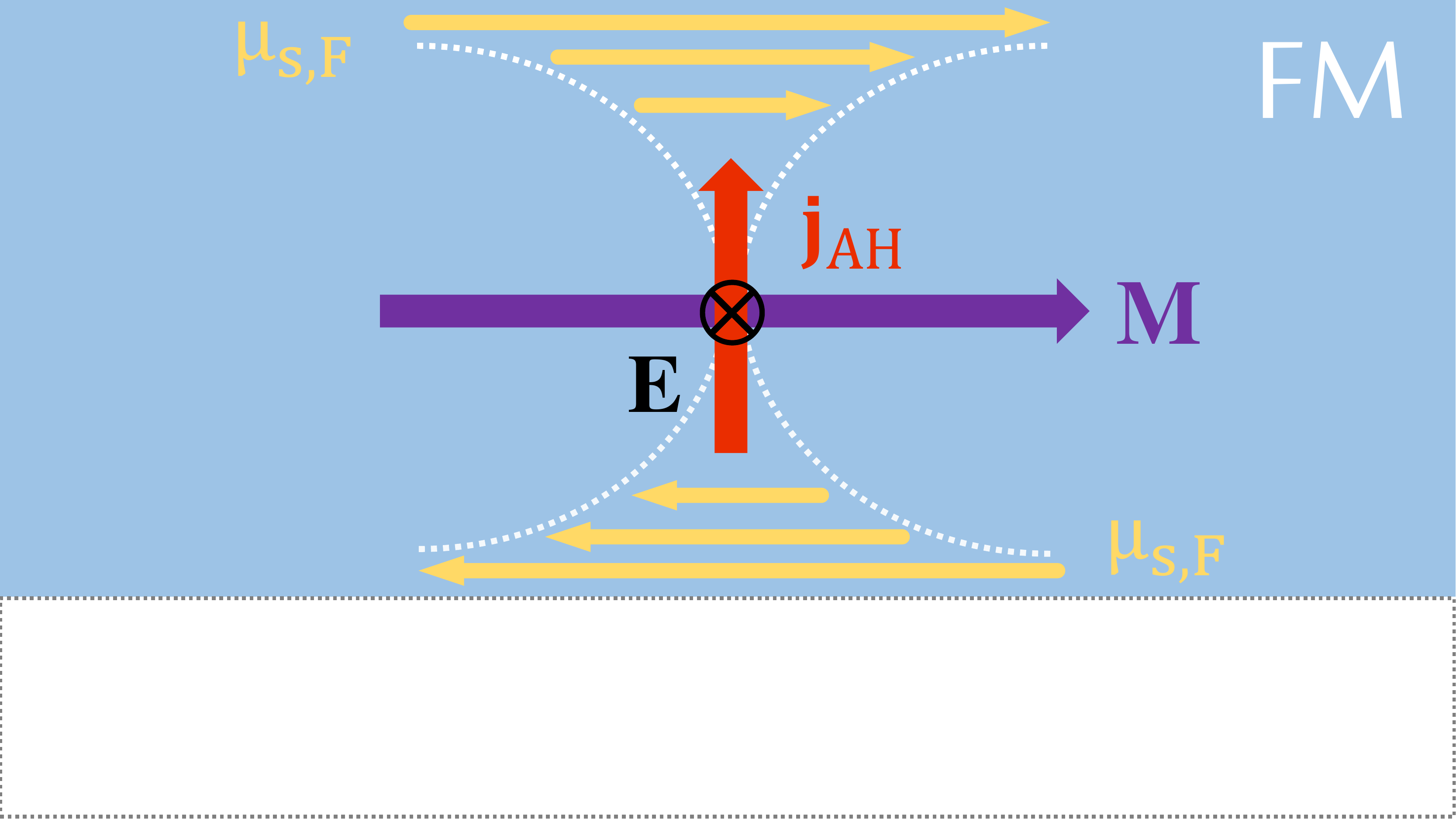}
\label{fig1a}}
\subfigure[]{
\includegraphics[width=.3\linewidth,trim={0 0 0 0}]{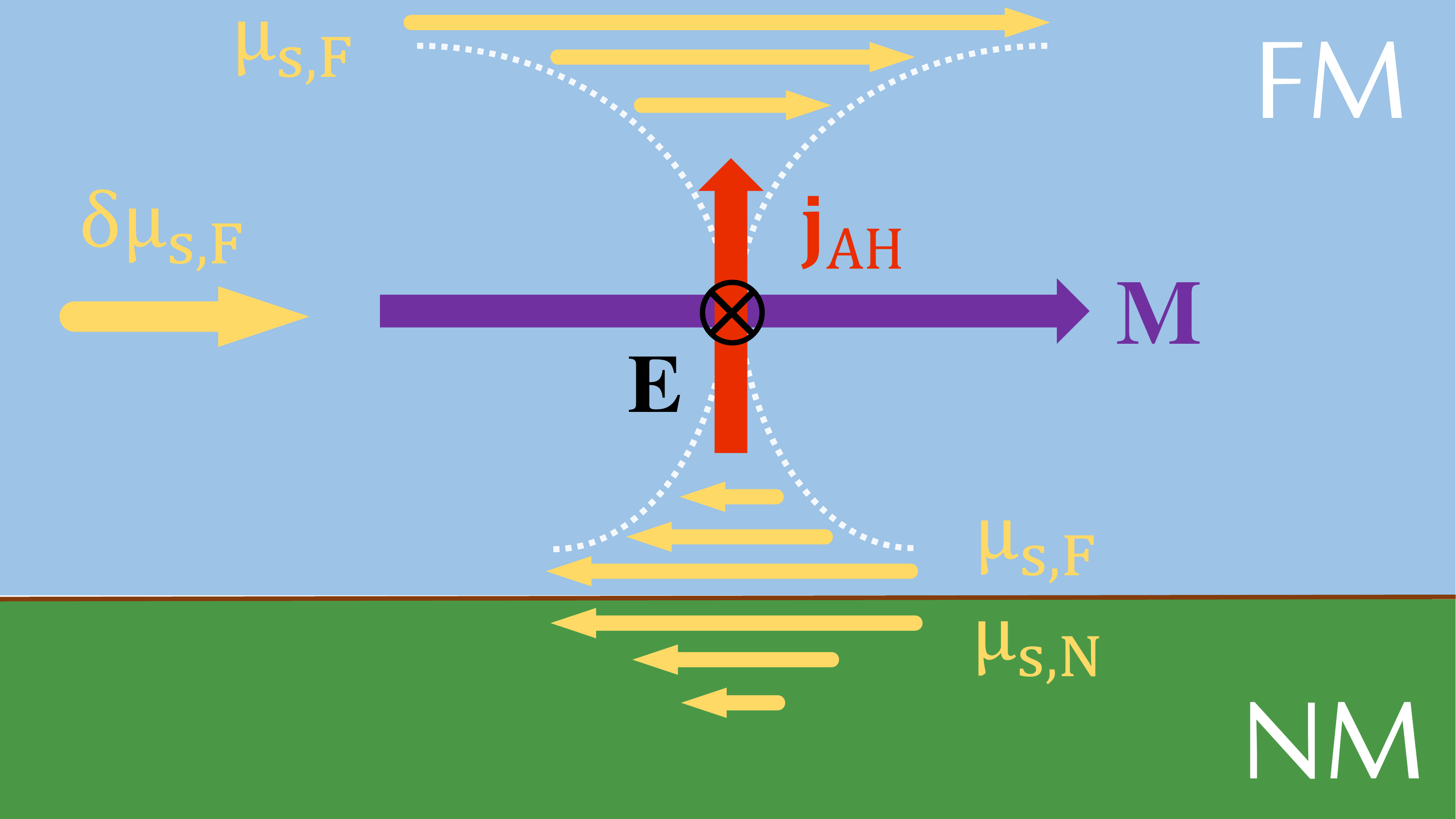}
\label{fig1b}}
\subfigure[]{
\includegraphics[width=.3\linewidth]{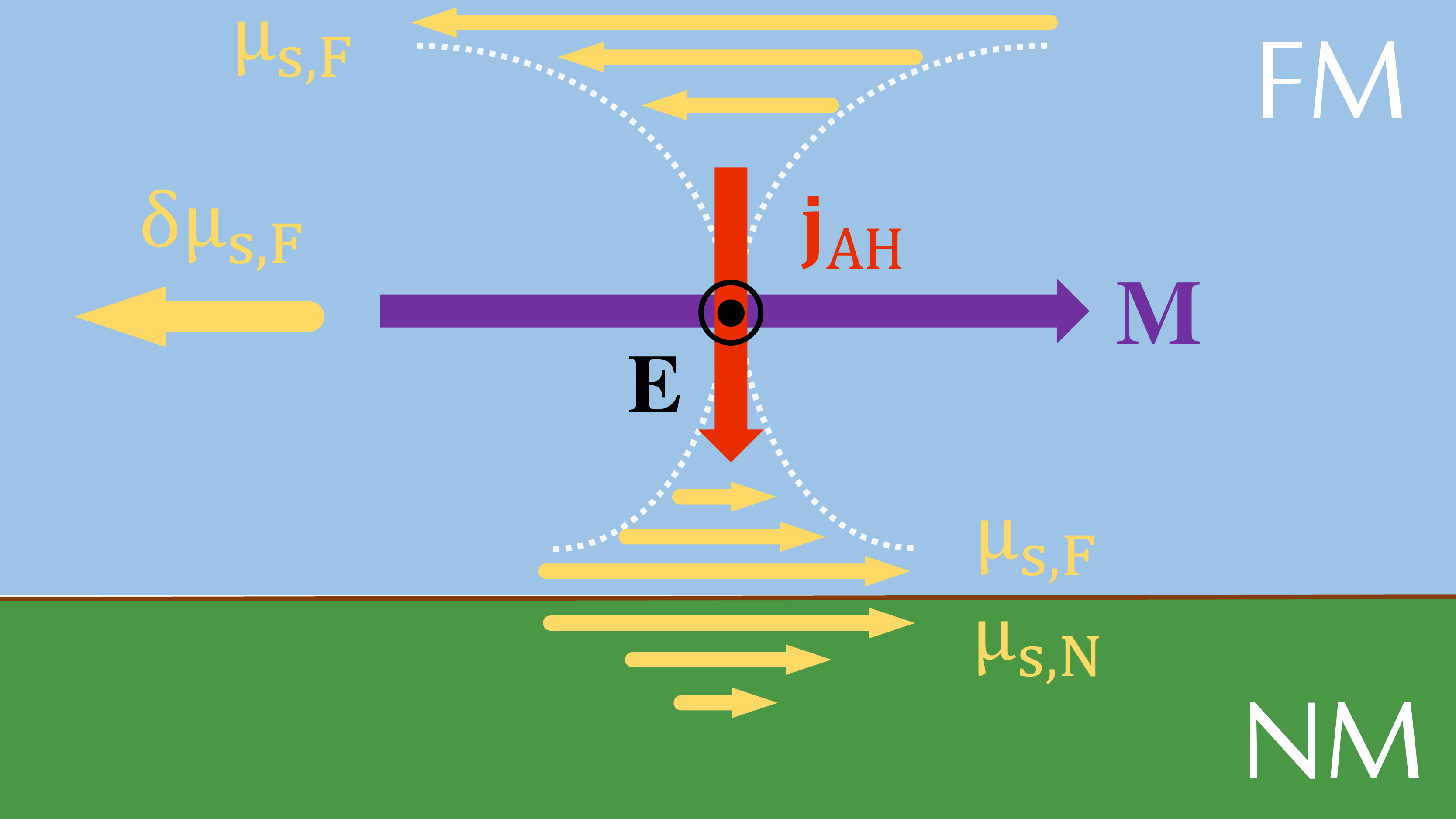}
\label{fig1c}}
\hspace*{\fill}
\caption{Schematics of the spin AH-UMR effect. (a) In a single ferromagnetic-metal (FM) layer, the spin accumulation $\mu_{s,F}$ (indicated by the small yellow arrows) has an antisymmetric distribution about the center line of the layer, with no net nonequilibrium spin density (spatially-averaged $\mu_{s,F}$) induced. So the overall spin AH-UMR is also zero. (b) The presence of a neighboring nonmagnetic-metal (NM) layer induces a net nonequilibrium spin density $\delta \mu_{s,F}$ (indicated by the large yellow arrow) in the FM layer, giving rise to a finite spin AH-UMR. (c) Reversing the electric field direction flips the direction of the AH current $\mathbf{j}_{\text{AH}}$ and thus the sign of $\delta \mu_{s,F}$, thereby changing the sign of the spin AH-UMR.}
\label{fig1}
\vspace*{-.5cm}
\end{figure*}

In what follows, we will first examine the coupled nonlinear transport of spin and charge induced by the spin AH effect in a ferromagnetic metal that is adjacent to a nonmagnetic-metal layer with negligible SH effect. Based on a generalized drift-diffusion model, an analytical expression of the UMR coefficient|a proper characterization of the nonlinear transport phenomenon|will be derived, which reveals the dependences of the spin AH-UMR effect on specific materials and geometric parameters of the bilayer. We will then generalize our results to bilayer structures comprised of a ferromagnetic metal and a heavy metal, wherein both the spin AH-UMR and USMR are present. And we predict, for particular choices of materials combinations, that the total UMR would exhibit a sign reversal when the thickness of either the ferromagnetic-metal or the heavy-metal layer is varied, which suggests a scheme to quantify the SH and spin AH angles experimentally through a UMR measurement. We will conclude with some materials considerations on both direct and indirect detections of the spin AH-UMR as well as the enhancement of the effect.

\section{Spin-dependent drift-diffusion model}

Consider a thin-film ferromagnetic layer located at $0<z<d_F$ placed on top of a nonmagnetic layer at $-d_N<z<0$, as shown in Figs.\,\ref{fig1}\textcolor{blue}{b,c}. Applying an in-plane electric field $\mathbf{E}$, the coupled drift-diffusion equations for charge and spin densities|taking into account the AH and the SH effects \cite{amin2019intrinsic} in the ferromagnetic layer|can be written as \cite{McGuire&Potter75IEEE_AMR,sZhang14JAP_AMR,kim2020generalized}
\begin{subequations}
\label{coupled}
\begin{align}
j_i
&=
\sigma 
\left(\mathcal{E}_i
+
p_{\sigma} \mathcal{E}^s_{ij}m_j \right)
-
\theta_{\text{SH}}^I \epsilon_{ijk} \mathcal{J}_{jk}
+
\theta_{\text{SH}}^A \epsilon_{ijk} m_j m_l\mathcal{J}_{kl}\,,
\\
\mathcal{J}_{ij}
&=
\sigma \left(\mathcal{E}^s_{ij} + p_{\sigma}\mathcal{E}_i m_j\right)
+
\theta_{\text{SH}}^I \epsilon_{ijk}j_k
+
\theta_{\text{SH}}^A \epsilon_{ilk} m_j m_l j_k\,,
\end{align}
\end{subequations}
where $j_i(z)$ and $\mathcal{J}_{ij}(z)$ are the local charge and spin current densities \footnote{By assuming spatial homogeneity in the $xy$ plane, all local quantities will have only a $z$ dependence.}, respectively, with the index $i$ indicating the direction of momentum flow and $j$ the spin polarization direction. $\theta_{\text{SH}}^I$ ($\theta_{\text{SH}}^A$) characterizes the strength of the isotropic (anisotropic) SH effect in the ferromagnet and $\mathbf{m}$ is a unit vector denoting the direction of the magnetization. In Eq.\,(\ref{coupled}), we have defined an effective local electric field felt by the electrons (in units with $e=1$) as $\bs{\mathcal{E}}(z) \equiv \mathbf{E} + \bs{\nabla}\mu_c(z)$, as well as an effective local `spin electric field', $\mathcal{E}^s_{ij}(z) \equiv \pd_i \mu_{s,j}(z)$. Here, $\mu_c$ and $\bs{\mu}_s$ are the charge and spin chemical potentials, respectively, whose gradients determine effective fields generated by local variations in the carrier densities \footnote{Note that, owing to the SH effect in the ferromagnet, the spin accumulation therein is, in general, in an arbitrary direction. This constitutes a generalization of the formalism presented in Ref.\,\cite{shulei2016prb}}. Furthermore, $\sigma(z)$ is the local conductivity and $p_{\sigma}$ the spin asymmetry in the linear conductivity in the ferromagnet.

To leading order in the Hall angles, the components of the spin current density polarized transverse to the magnetization|as well as the relevant boundary conditions|are decoupled from the charge current density~\cite{kim2020generalized}, the details of which are presented in Appendix \ref{appendix_a}. Thus, Eq.\,(\ref{coupled}) reduces to
\begin{subequations}
\label{coupled_vectorial}
\begin{align}
\mathbf{j}
&=
\sigma \mathbf{E} + \sigma_0 \left(\bs{\nabla}\mu_c + p_{\sigma} \bs{\nabla} \mu_s + p_{\sigma}\theta_{\text{SAH}} \mathbf{m} \times \mathbf{E}\right),
\\
\bs{\mathcal{J}}
&=
p_{\sigma}\sigma \mathbf{E} + \sigma_0 \left(p_{\sigma} \bs{\nabla}\mu_c + \bs{\nabla} \mu_s + \theta_{\text{SAH}} \mathbf{m} \times \mathbf{E}\right),
\end{align}
\end{subequations}
where $\sigma_0$ is the equilibrium conductivity, $\theta_{\text{SAH}} (\equiv \theta_{\text{SH}}^I + \theta_{\text{SH}}^A$) is the spin AH angle with $p_{\sigma} \theta_{\text{SAH}}$ being its charge counterpart, $\mathcal{J}_i \equiv \mathcal{J}_{ij}m_j$ and $\mu_s \equiv \bs{\mu}_s \cdot \mathbf{m}$ are the longitudinal components of the spin current density and spin chemical potential, respectively.

The local conductivity in Eq.\,(\ref{coupled_vectorial}) may be expressed as the sum of the conductivities of each spin channel, given by
\begin{equation}
\label{sigma_z}
\sigma^{\alpha}(z)=\nu^{\alpha}\left[n_0^{\alpha} + n^{\alpha}(z)\right]\,,
\end{equation}
where $\alpha=+(-)$ denotes the spin moment parallel (antiparallel) to the local magnetization, $n_0^{\alpha}$ and $\nu^{\alpha}$ are the equilibrium electron density and mobility of spin-$\alpha$ electrons, respectively, which in combination give rise to the longitudinal conductivity in the linear response regime ( \textit{i.e.}, $\sigma_0=\sum_{\alpha} \sigma_0^{\alpha}~\text{with}~ \sigma_0^{\alpha}=\nu^{\alpha}n_0^{\alpha}$) as well as the spin asymmetry in the conductivity by $p_{\sigma}\equiv (\sigma_0^+ - \sigma_0^-)/(\sigma_0^+ + \sigma_0^-)$. And $n^{\alpha}(z)$ are the current-induced nonequilibrium carrier densities that are responsible for the UMR. Local charge neutrality is assumed, \textit{i.e.}, $\sum_{\alpha}n^{\alpha}(z)=0$, for the metallic system.

The charge and spin chemical potentials may also be defined as $\mu_c \equiv (\mu^+ + \mu^-)/2$ and $\mu_s \equiv (\mu^+ - \mu^-)/2$, where $\mu^{\alpha}(z)$ is the spin-dependent chemical potential parallel to the magnetization, which is related to the nonequilibrium electron density through
\begin{equation}
\label{mu(z)}
\mu^{\alpha}(z)=\left[N^{\alpha}(\epsilon_F)\right]^{-1} n^{\alpha}(z) - \phi(z)\,,
\end{equation}
with $N^{\alpha}(\epsilon_F)$ the density of states of spin-$\alpha$ electrons at the Fermi level and $\phi(z)$ the spin-independent part of the chemical potential.

In the presence of spin-flip scattering, the charge- and spin-current densities satisfy, respectively, the following continuity 
equations~\cite{valet1993prb}
\begin{equation}
\label{cont}
\bs{\nabla} \cdot \mathbf{j}=
0~~\text{and}~~\bs{\nabla} \cdot \bs{\mathcal{J}}=
\frac{2}{\tau_{sf}} n_s\,,
\end{equation}
where $\tau_{sf}$ is the spin-flip relaxation time and $n_s(z)=\left(1-p_N^2\right)N\left(\epsilon_F\right)\mu_s(z)$ is the local spin density at the Fermi level, with $N(\epsilon_F)=\sum_{\alpha} N^{\alpha}(\epsilon_F)$ the total density of states and $p_N \equiv (N^{+}-N^{-})/(N^{+} +N^{-})$.

Inserting Eq.\,(\ref{coupled_vectorial}) into Eq.\,(\ref{cont}), we obtain a set of differential equations for the charge and spin chemical potentials
\begin{subequations}
\label{decomposed_cont_fm}
\begin{align}
\label{sym_cont_fm}
&\frac{d^2}{dz^2}\mu_{c,F}(z) + p_{\sigma}\frac{d^2}{dz^2}\mu_{s,F}(z)=0\,,
\\
\label{antisym_cont_fm}
&\frac{d^2}{dz^2}\mu_{s,F}(z) - \frac{\mu_{s,F}(z)}{\lambda_F^2}=0\,,
\end{align}
\end{subequations}
where $\lambda_F=\sqrt{\sigma_{0,F}(1-p_{\sigma}^2)\tau_{sf}/2N_F(\epsilon_F)(1-p_N^2)}$ is the spin diffusion length of the ferromagnetic metal.

In a nonmagnetic metal, where $p_{\sigma},p_N=0$, the charge and spin chemical potentials satisfy
\begin{subequations}
\label{decomposed_cont_nm}
\begin{align}
\label{sym_cont_nm}
&\frac{d^2}{dz^2}\mu_{c,N}(z) =0\,,
\\
\label{antisym_cont_nm}
&\frac{d^2}{dz^2}\mu_{s,N}(z) - \frac{\mu_{s,N}(z)}{\lambda_N^2}=0\,,
\end{align}
\end{subequations}
where $\lambda_N=\sqrt{\tau_{sf}\sigma_{0,N}/2N_N(\epsilon_F)}$ is the spin diffusion length of the normal metal with $N_N(\epsilon_F)$ the density of states of electrons at the Fermi level.

At the interface of the bilayer ($z=0$), we assume both the current density and chemical potential for each conduction channel are continuous (see Appendix \ref{appendix_b} for a more detailed discussion on the boundary conditions), \textit{i.e.}, $\mu^{\alpha}_N(0^-)=\mu^{\alpha}_F(0^+)$ and $j_z^{\alpha}(0^-)=j_z^{\alpha}(0^+)$. And open boundary conditions are imposed at the two outer surfaces, \textit{i.e.}, $j_z^{\alpha}(-d_N)=j_z^{\alpha}(d_F)=0$. 
\begin{figure}[b]
\centering 
\hspace*{-1.0cm}
\subfigure{\label{fig2a}
\includegraphics[width=0.85\linewidth]{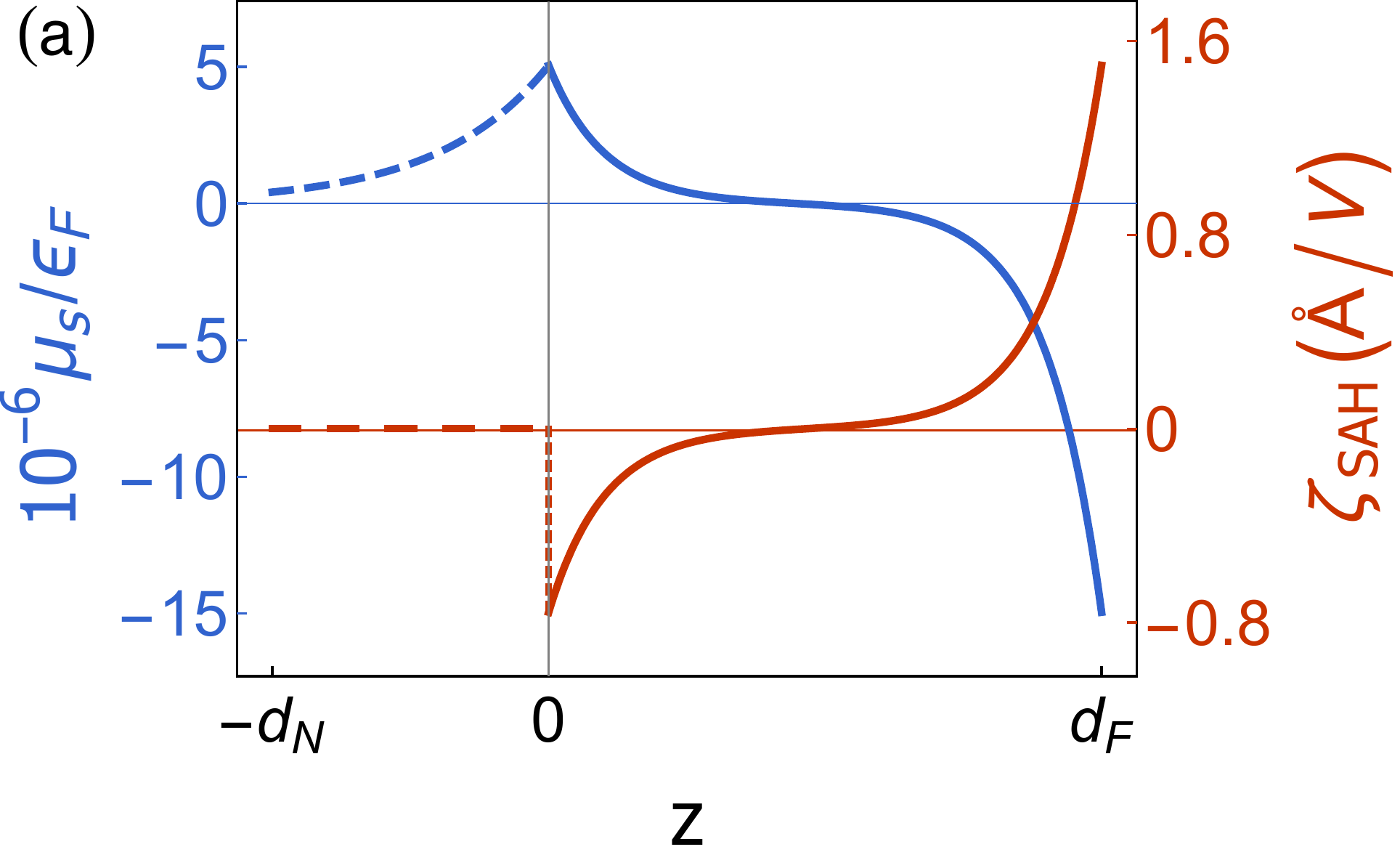}}
\subfigure{
\includegraphics[width=0.85\linewidth]{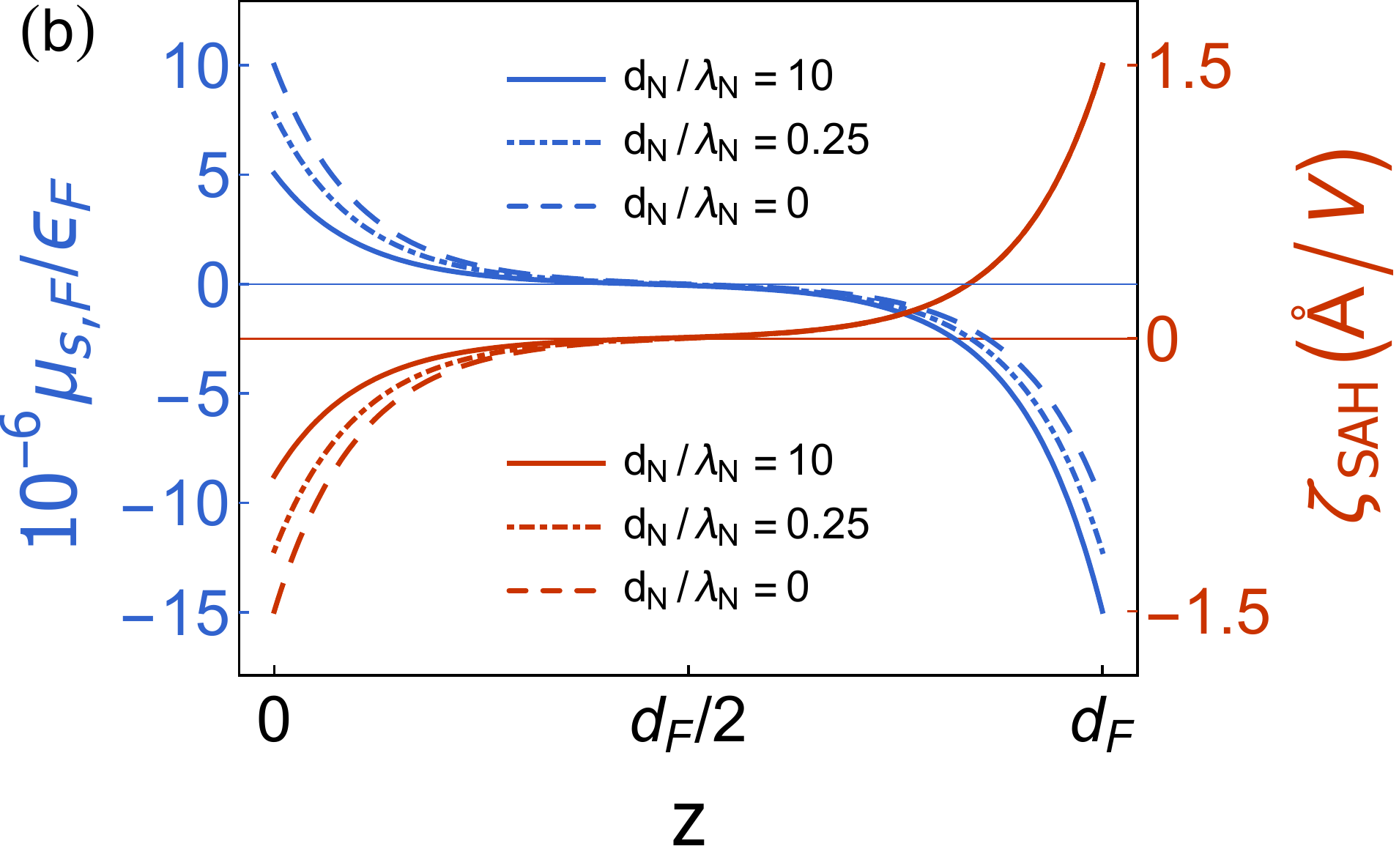}
\label{fig2b}}
\hspace*{\fill}
\caption{Spatial dependences of the spin accumulation and spin AH-UMR coefficient in the bilayer system. (a) z-dependences of the spin accumulation $\mu_s$ and spin AH-UMR coefficient $\zeta_{\text{SAH}}$ in the FM and NM layers. Note the absence of the spin AH-UMR in the NM. (b) Plots of the spin accumulation and spin AH-UMR coefficient in the FM for different thicknesses of the NM layer. Parameters used: $\lambda_F=10$~nm, $\lambda_N=20$~nm, $\theta_{\text{SAH}}=0.05$, $\epsilon_F=5$~eV, $\sigma_{0,F}=\sigma_{0,N}=0.033$ (\textmu\textOmega\;cm)$^{-1}$, $p_{\sigma}=0.7$, and $p_N=0.2$.} 
\label{fig2}
\end{figure}

\section{Spin accumulation and nonlinear transport}

Without loss of generality, let us fix the electric field $\mathbf{E}$ in the $x$ direction, $\mathbf{E}=E_x \mathbf{x}$, and set the magnetization to point in the 
$y$ direction|in which case the 
magnitude of the longitudinal UMR would reach the maximum. The total charge current density, given by $\sum_{\alpha}j^{\alpha}(z)$, can be divided into two parts: $\mathbf{j}=\mathbf{j}^{(1)}+\mathbf{j}^{(2)}$, with a linear component $\mathbf{j}^{(1)}$ that is proportional to $E_x$ and a nonlinear one that is quadratic in $E_x$. The latter can be expressed as  
\begin{equation}\label{eq:j^(2)}
    j_x^{(2)}(z)=(\nu^{+}-\nu^{-})\mu_s(z)\bar{\mathcal{H}}(N^{+},N^{-})E_x
\end{equation}
where $\bar{\mathcal{H}}(N^{+},N^{-})$ is the harmonic mean of the density of states. And the spin accumulation $\mu_s$ is linear in $\theta_{\text{SAH}}E_x$, resulting in $j_x^{(2)}\propto E_x^2$. Note that $j_x^{(2)}$ only emerges in the ferromagnetic layer wherein $\nu^{+}\neq \nu^{-}$. 

In order to properly quantify the nonlinear charge current, we introduce a UMR coefficient $\zeta(z)$ as
\begin{equation}\label{eq:zeta1}
\zeta(z)
\equiv
\frac{\sigma _{xx}(z,E_{x})-\sigma
_{xx}(z,-E_{x})}{\sigma_{0}E_{x}}\;,
\end{equation} 
where $\sigma_{ij}=j_i/E_j$ is the conductivity tensor, and $\sigma_{0}$ is the linear longitudinal conductivity. The UMR coefficient $\zeta$ is so defined that its magnitude is independent of the electric field. The dimension of $\zeta$ is length per Volt, the inverse of which sets the scale of the electric field for which the nonlinear longitudinal conductivity|given by $j_{x}^{(2)}/E_x$|becomes comparable to its linear counterpart. 

The spatial distribution of the spin AH-UMR coefficient $\zeta_{\text{SAH}}$ is displayed in Fig.~\ref{fig2}\textcolor{blue}{a}, along with that of the spin chemical potential $\mu_s$ (or the spin accumulation) in the bilayer structure. There is a clear correlation between the two quantities: $\zeta_{\text{SAH}}$ goes to zero wherever $\mu_s$ vanishes. Furthermore, the spin AH-UMR completely comes from the ferromagnetic layer, in which the electron mobility is spin-dependent. Within the nonmagnetic layer, for which $\nu^{+} = \nu^{-}$, $\zeta_{\text{SAH}}$ vanishes everywhere despite the remnant $\mu_s$ near the interface. These observations are in full agreement with Eq.\,(\ref{eq:j^(2)}).

Although the nonmagnetic layer in question neither plays an active role as a spin polarizer nor accommodates any nonlinear charge transport, it is still indispensable to the generation of a net spin AH-UMR in its neighboring ferromagnetic layer. In the absence of the nonmagnetic layer, the spin accumulation $\mu_s$ and thus the local spin AH-UMR coefficient $\zeta_{\text{SAH}}$ have antisymmetric distributions about the center line of the ferromagnetic layer, as shown by the dashed lines in Fig.\,\ref{fig2}\textcolor{blue}{b}. In this case, the total (spatially-averaged) spin AH-UMR is zero, as a result of the lack of a net nonequilibrium spin density. 

From a symmetry perspective, a net current-induced spin density is allowed, when and only when a system lacks inversion symmetry. For the present case, the nonmagnetic layer introduces structural inversion asymmetry, and makes a net nonequilibrium spin density achievable in the ferromagnetic layer next to it. Physically, it ``absorbs" spin accumulation at the interface from the ferromagnetic layer, leaving a net nonequilibrium spin density in the latter, as illustrated by the dash-dotted and solid lines in Fig.\,\ref{fig2}\textcolor{blue}{b}.

\section{Spatially-averaged spin AH-UMR}

By taking the spatial average of the overall UMR coefficient over the thickness of the bilayer, $\bar{\zeta}\equiv \int_{-d_N}^{d_F} dz \;\zeta(z)/(d_N+d_F)$, we find that, up to $\mathcal{O}(\theta_{\text{SAH}})$, the spatially-averaged spin AH-UMR coefficient reads
\begin{equation}
\label{SAH-UMR}
\bar{\zeta}_{\text{SAH}}
=p_F \theta _{\text{SAH}}\left(\frac{\lambda_F}{\epsilon_F}\right)\mathcal{G}\left(\frac{d_{F}}{\lambda_{F}},\frac{d_{N}}{\lambda_{N}};\frac{\sigma_{0,F}}{\sigma_{0,N}},\frac{\lambda_{F}}{\lambda_{N}}\right)\,,
\end{equation}
where $p_F(=p_{\sigma} - p_N)$ characterizes the overall spin asymmetry of electron mobility for the ferromagnetic layer, and the thickness dependence of the spin AH-UMR is encapsulated in the dimensionless $\mathcal{G}$ function as
\begin{equation}
\label{eq:G}
\mathcal{G}\left(s,t;u,v \right)=\frac{3\left(\frac{uv}{(uv)\cdot s + t}\right)\tanh(s)\tanh(\frac{s}{2})}
{1+ \left(1-p_{\sigma}^2\right)\left(\frac{u}{v}\right) \tanh(s) \coth (t)}.
\end{equation}
For simplicity, we have adopted the free-electron model whereby  $N_F^{\alpha}=3n_{0,F}^{\alpha}/2\epsilon_F$ with $\epsilon_F$ the Fermi energy of conduction electrons in the ferromagnet. Equations\,(\ref{SAH-UMR}) and (\ref{eq:G}) are the main results of the paper. 

Several remarks regarding the spin AH-UMR are in order. 

1) The spin AH-UMR coefficient, to leading order, is proportional to the spin AH angle $\theta_{\text{SAH}}$, in contrast to the USMR effect, which is proportional to the SH angle of the heavy-metal layer~\cite{shulei2016prb}.

2) The spin AH-UMR coefficient is also linear in $p_F$, as is the USMR~\cite{shulei2016prb}. This is not surprising, as the conversion of a net nonequilibrium spin density to a (nonlinear) charge current relies entirely on the spin asymmetry in electron scatterings. 

3) The ratio $\frac{\lambda_F}{\epsilon_F}$ has the same dimension as the spin AH-UMR coefficient (with $e=1$). In fact, the prefactor of the averaged UMR coefficient|the thickness independent part in Eq.\,(\ref{SAH-UMR})|sets the scale of the maximum spin AH-UMR that one can obtain for a given ferromagnet. For a typical transition metal with $p_F=0.3$, $\theta_{\text{SAH}}=0.02$, $\lambda_F=100$~nm, and $\epsilon_F=5$~eV, the upper bound of the spin AH-UMR coefficient, $\bar{\zeta}_{\text{SAH}}$, is of the order of 1\text{~\AA/V}.  

\begin{figure}[htb]
\includegraphics[width=.45\textwidth]{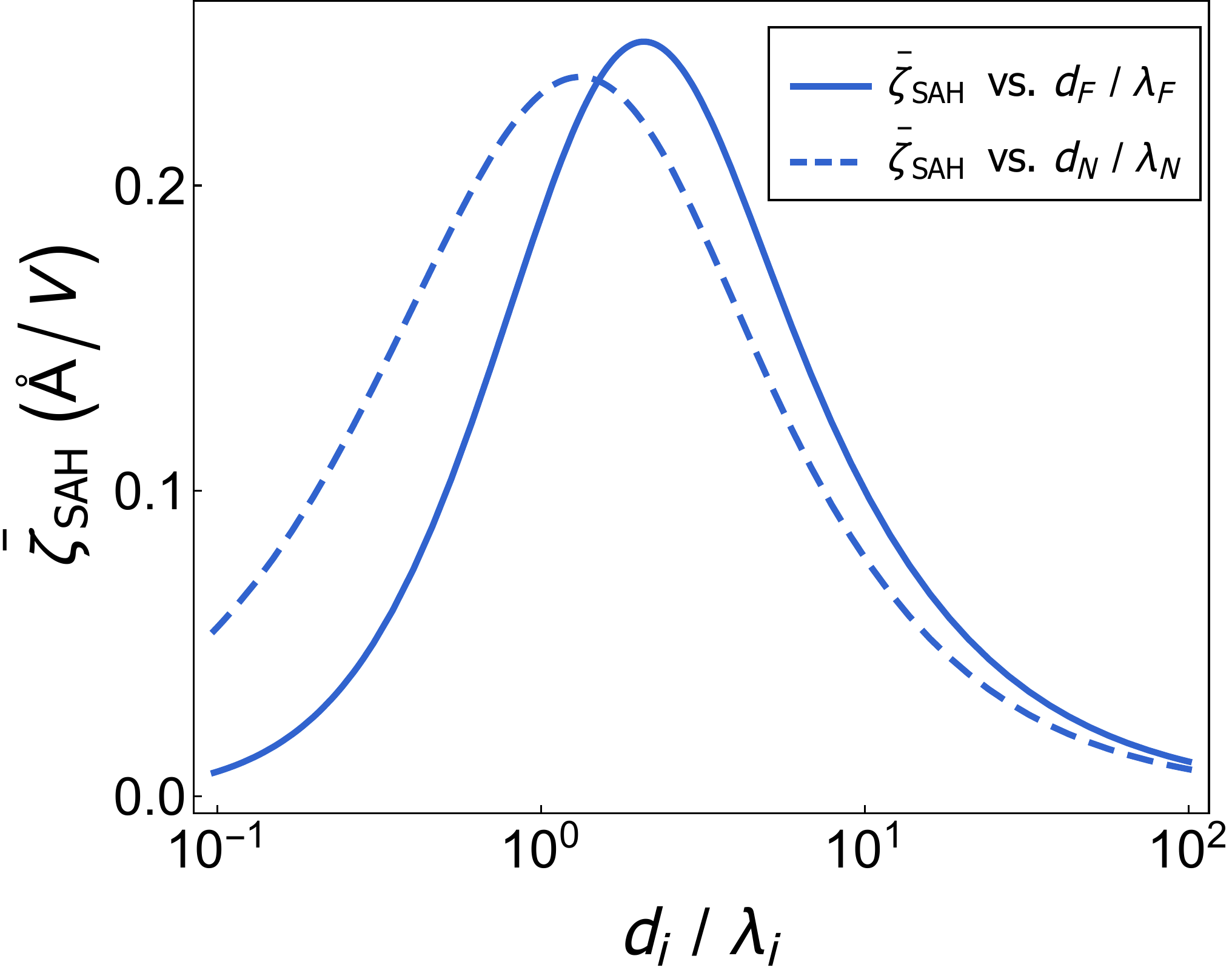}
\caption{Thickness dependence of the spin AH-UMR coefficient. Dependence of $\bar{\zeta}_{\text{SAH}}$ on the thickness ($i=F,N$) of the FM layer for $d_N=15$ nm (solid) and on the thickness of the NM layer for $d_F=15$ nm (dashed). See the caption of Fig.\,\ref{fig2} for the list of materials parameters used.}
\label{fig3}
\end{figure}

4) Information about how other geometric and materials parameters of a magnetic bilayer would shape the spin AH-UMR is all encoded in the dimensionless $\mathcal{G}$ function given by Eq.\,(\ref{eq:G}). The first two variables, $d_F/\lambda_F$ and $d_N/\lambda_N$, indicate that the dependences of the spin AH-UMR on the thicknesses of the ferromagnetic and nonmagnetic layers must scale with their respective spin diffusion lengths, as plotted in Fig.\,\ref{fig3}. 

5) Another remarkable property of the spin AH-UMR|revealed by the $\mathcal{G}$ function|is that it increases monotonically with the ratio $\frac{\lambda_F}{\lambda_N}$, which would be useful for guiding the search for magnetic bilayers with a sizable spin AH-UMR effect.  

\section{UMR sign reversal and SH/spin-AH angle quantification}

In a magnetic bilayer consisting of a ferromagnetic metal and a heavy metal, both spin AH and SH effects, in principle, may contribute to the total UMR measured in the bilayer. And their contributions turn out to be additive, \textit{i.e.}, $\bar{\zeta}=\bar{\zeta}_{\text{SAH}}+\bar{\zeta}_{\text{SH}}$ (see Appendix \ref{appendix_b} for the full expression of $\bar{\zeta}$). The ratio of the two UMR contributions due to spin-dependent scattering|provided electron-magnon scattering is suppressed by applying a magnetic field or lowering the temperature~\cite{avci2018prl}|takes a rather neat form 
\begin{equation}
\frac{\bar{\zeta}_{\text{SAH}}}{\bar{\zeta}_{\text{SH}}}
=
\frac{\theta_{\text{SAH}} \lambda_F \tanh\left(\frac{d_F}{2\lambda_F}\right)}{\theta_{\text{SH}} \lambda_N \tanh \left(\frac{d_N}{2\lambda_N}\right)}\,.
\end{equation}
It is worthy to note that this ratio depends on only a few parameters, namely the SH/spin-AH angle, the spin diffusion length, and the thickness of each layer. 

\begin{figure}[t]
\includegraphics[width=.45\textwidth]{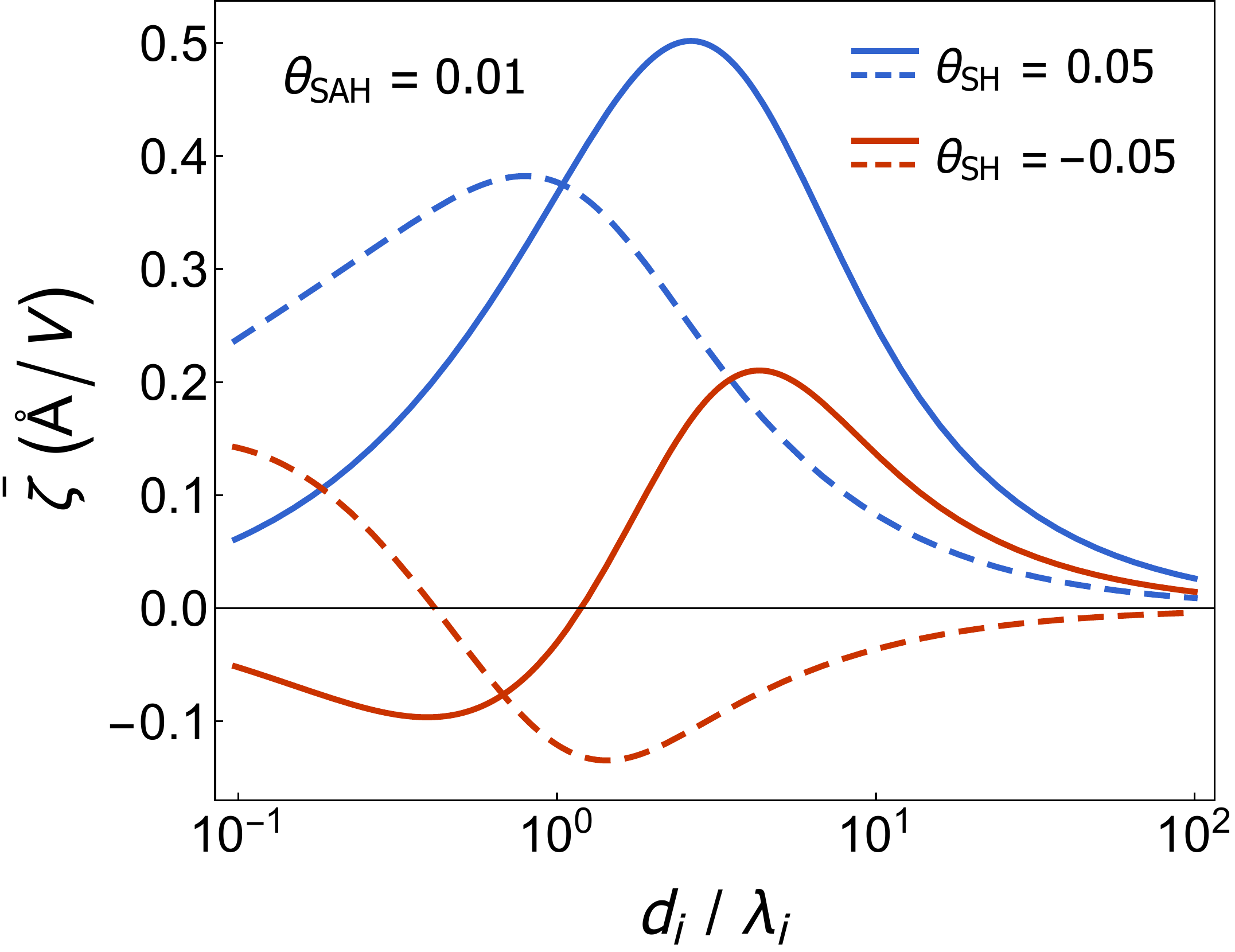}
\caption{Thickness dependence of the total UMR coefficient. Dependence of $\bar{\zeta}$ on the thickness ($i=F,N$) of the FM layer for $d_N=5$ nm (solid) and on the thickness of the NM layer for $d_F=10$ nm (dashed). Other parameters used: $\lambda_F=20$ nm, $\lambda_N=5$ nm,  $\epsilon_F=5$ eV, $\sigma_{0,F}=\sigma_{0,N}=0.033$ (\textmu\textOmega\;cm)$^{-1}$, $p_{\sigma}=0.7$, and $p_N=0.2$.}
\label{fig4:sign}
\end{figure}

The simple relation between $\bar{\zeta}_{\text{SAH}}$ and $\bar{\zeta}_{\text{SH}}$ nonetheless has a remarkable physical consequence: the total UMR coefficient of such a magnetic bilayer may exhibit qualitatively different thickness dependences, depending on the relative signs of the spin AH and SH angles. When $\theta_{\text{SAH}}$ and $\theta_{\text{SH}}$ have the same sign, the associated contributions simply add up (see the blue curves in Fig.~\ref{fig4:sign}). It becomes more intriguing when $\theta_{\text{SAH}}$ and $\theta_{\text{SH}}$ are of opposite signs. In this case, the total UMR inevitably undergoes a sign reversal as the thickness of either layer is varied (see the red curves in Fig.\,\ref{fig4:sign}). 

At the sign reversal point, the ratio of the Hall angles fulfills the following condition:
\begin{equation}
\frac{\theta_{\text{SH}}}{\theta_{\text{SAH}}}=-
\frac{\lambda_F \tanh\left(\frac{d_F}{2\lambda_F}\right)}{ \lambda_N \tanh \left(\frac{d_N}{2\lambda_N}\right)}\,,
\end{equation}
which can be used to experimentally quantify the SH (spin AH) angle of the heavy-metal (ferromagnetic-metal) layer, provided the Hall angle and spin diffusion length of the other layer|which serves as a reference layer|are known.

\section{Proposal for detecting spin AH-UMR}

As compared to the USMR, there are more options of materials systems for probing the spin AH-UMR effect. For the USMR, the nonmagnetic heavy-metal layer plays a central role in creating nonequilibrium spin density|via the SH effect|in the adjacent ferromagnetic layer. But this is not the case for the spin AH-UMR. For the spin AH-UMR, it is the ferromagnetic layer that serves as the spin polarizer and, hence, the choice of its neighboring layer is not necessarily limited to nonmagnetic materials with a strong SH effect. 

For instance, the spin AH-UMR effect, in principle, can also be hosted in bilayers comprised of a ferromagnetic metal and a normal metal, such as Cu, Al, or Ag. Given the weak SH effect in the normal-metal layer, the spin AH-UMR is expected to dominate over the USMR in these systems, making the detection of the former more straightforward. There is, perhaps, also a downside to these metallic bilayers|the spin AH-UMR therein is likely to be much smaller than that in magnetic bilayers with heavy-metals, as normal metals with a weak SH effect are oftentimes also poor ``spin sinks" with long spin diffusion lengths~\cite{Bass07JP-CM_spin-diffusion}, which would diminish the spin AH-UMR effect (especially when the ratio $\frac{\lambda_F}{\lambda_N}$ is small~\footnote{For instance, a UMR was not detected in Cu$|$Co bilayers~\cite{Gambardella15APL_MR-HL} (or, at least, was demonstrated to be much smaller than that in W$|$Co bilayers), which is not surprising, as the $\frac{\lambda_F}{\lambda_N}$ ratio for the former is about two orders of magnitude smaller than that for the latter.}, as was discussed in a previous section). 

The shortcoming of normal metals with long spin diffusion lengths may be compensated for by choosing a ferromagnetic layer with low carrier density. To see this, let us insert Eq.\,(\ref{eq:j^(2)}) together with $\sigma_0=\sum_{\alpha} n_0^{\alpha} \nu^{\alpha}$ into Eq.\,(\ref{eq:zeta1}), which yields 
\begin{equation}
\label{eq:zeta2}
    \zeta_{\text{SAH}} \propto \left(\frac{\nu^{+}-\nu^{-}}{\nu^{+}+\nu^{-}}\right) \frac{\theta_{\text{SAH}}}{n_0}\,.
\end{equation}
The above relation conveys a valuable piece of information: the lower the equilibrium carrier density of the ferromagnet, the larger the UMR coefficient. Thus, a more sizable spin AH-UMR is expected to arise in bilayers consisting of a normal-metal and a ferromagnetic semiconductor [\textit{e.g.}, (Ga,Mn)As]~\cite{Jungwirth15PRB_UMR-DMS} whose carrier density is usually two to three orders of magnitude smaller than that of a ferromagnetic metal. 

In magnetic bilayers comprised of nonmagnetic materials with strong spin-orbit coupling, the coexistence of the spin AH-UMR and USMR poses a challenge to differentiate the two. But, interestingly, they may also conspire to bring about a sign reversal of the overall UMR when the thickness of either layer is varied, a distinct transport signature that would not appear when either effect stands alone. This can be experimentally verified by contrasting the thickness dependences of the total UMR in two ferromagnetic-metal$|$heavy-metal bilayers, either with  different heavy-metal layers whose SH angles are of opposite signs (such Pt and $\beta$-Ta~\cite{liu2012spin}) or with different ferromagnetic-metal layers whose spin AH angles are of opposite signs (such as Fe and Gd~\cite{xxZhang10EPL_AH_Fe-Gd}). We anticipate that the results of such comparative measurements will resemble what are shown in Fig.\,\ref{fig4:sign}, with one bearing a sign change and the other not. On a related note, a sign change of the UMR was recently observed in single-crystalline Fe$|$Pt bilayers as the thickness of the Fe layer was increased~\cite{yWu21PRB_UMR-sign-revs}, implying possible competition between the spin AH-UMR and the USMR.

Other nonlinear effects that may intertwine with the spin AH-UMR are the anomalous Nernst~\cite{weiler2012local} and spin Seebeck~\cite{kikkawa2013longitudinal} effects: the Joule heating may induce a vertical temperature gradient across the ferromagnetic layer, which would in turn give rise to a nonlinear current in the direction of $\mathbf{m}\times \bs{\nabla}T$. However, one can separate the UMR contribution from the thermal contribution based on their different dependences on the direction of the applied electric field: the former is proportional to $\hat{\mathbf{z}}\cdot(\mathbf{m}\times \mathbf{E})$, whereas the latter is independent of the relative orientation between  $\mathbf{E}$ and $\mathbf{m}$, as the temperature gradient only relies on the magnitude of electric field (\textit{i.e.}, $\bs{\nabla}T \propto \lvert \mathbf{E} \rvert^2$).

\section*{Acknowledgments}

We thank Chong Bi, Kirill Belashchenko, and Xin Fan for helpful discussions. This work was supported by the College of Arts and Sciences, Case Western Reserve University. \\

\appendix 

\section{Decoupling of Transverse Spin Chemical Potentials}
\label{appendix_a}

Recall Eq.\,(\ref{coupled}) in the main text
\begin{subequations}
\label{SM_coupled}
\begin{align}
j_i
&=
\sigma 
\left(\mathcal{E}_i
+
p_{\sigma} \mathcal{E}^s_{ij}m_j \right)
-
\theta_{\text{SH}}^I \epsilon_{ijk} \mathcal{J}_{jk}
+
\theta_{\text{SH}}^A \epsilon_{ijk} m_j m_l\mathcal{J}_{kl}\,,
\\
\mathcal{J}_{ij}
&=
\sigma \left(\mathcal{E}^s_{ij} + p_{\sigma}\mathcal{E}_i m_j\right)
+
\theta_{\text{SH}}^I \epsilon_{ijk}j_k
+
\theta_{\text{SH}}^A \epsilon_{ilk} m_j m_l j_k\,.
\end{align}
\end{subequations}
Here, $\mathcal{J}_{ij}$ is the spin current density tensor, in which the index $i$ indicates the direction of electron momentum flow and $j$ the spin polarization direction. Let us introduce the following decomposition of the spin current density
\begin{equation}
\mathcal{J}_{ij}
=
\mathcal{J}_i m_j + \mathcal{J}_i^{\perp_1} p_j + \mathcal{J}_i^{\perp_2} q_j\,,
\end{equation}
where $\mathbf{m}$, $\mathbf{p}$ and $\mathbf{q}$ are three mutually orthogonal unit vectors. $\mathcal{J}_i \equiv \mathcal{J}_{ij}m_j$ is the spin current density vector with the spins polarized along the magnetization, while $\mathcal{J}_i^{\perp_1} \equiv \mathcal{J}_{ij}p_j$ and $\mathcal{J}_i^{\perp_2} \equiv \mathcal{J}_{ij}q_j$ are two spin current density vectors with spins polarized perpendicular to the magnetization and to each other. Using the definitions of the effective local electric and spin electric fields, $\bs{\mathcal{E}}\equiv \mathbf{E} + \bs{\nabla} \mu_c$ and $\mathcal{E}^s_{ij}\equiv \pd_i \mu_{s,j}$, to leading order in the Hall angles, the charge and spin current density vectors read
\begin{subequations}
\label{SM_coupled_vectorial}
\begin{align}
&\mathbf{j}
=
\sigma \mathbf{E} + \sigma_0 \left(\bs{\nabla}\mu_c + p_{\sigma} \bs{\nabla} \mu_s + p_{\sigma}\theta_{\text{SAH}} \mathbf{m} \times \mathbf{E}\right),
\\
&\bs{\mathcal{J}}
=
p_{\sigma}\sigma \mathbf{E} + \sigma_0 \left(p_{\sigma} \bs{\nabla}\mu_c + \bs{\nabla} \mu_s + \theta_{\text{SAH}} \mathbf{m} \times \mathbf{E}\right),
\\
&\bs{\mathcal{J}}^{\perp_1}
=
\sigma_0 \left(\bs{\nabla} \mu_s^{\perp_1} + \theta_{\text{SH}}^I  \mathbf{p} \times \mathbf{E}\right),
\\
&\bs{\mathcal{J}}^{\perp_2}
=
\sigma_0 \left(\bs{\nabla} \mu_s^{\perp_2} + \theta_{\text{SH}}^I  \mathbf{q} \times \mathbf{E}\right),
\end{align}
\end{subequations}
where $\theta_{\text{SAH}} = \theta_{\text{SH}}^I + \theta_{\text{SH}}^A$ is the spin AH angle, $\mu_s^{\perp_1} \equiv \bs{\mu}_s \cdot \mathbf{p}$ and $\mu_s^{\perp_2} \equiv \bs{\mu}_s \cdot \mathbf{q}$.

In the presence of spin-flip scattering, and assuming local charge neutrality, the charge and spin current densities obey the following continuity equations
\begin{subequations}
\label{SM_cont}
\begin{align}
\bs{\nabla} \cdot \mathbf{j}
&=
0\,,
\\
\pd_i \mathcal{J}_{ij}
&=
\frac{2}{\tau_{sf}} n_{s,j}\,,
\end{align}
\end{subequations}
where $\tau_{sf}$ is the spin-flip relaxation time and $\mathbf{n}_s(z)=\left(1-p_N^2\right)N\left(\epsilon_F\right)\bs{\mu}_s(z)$ is the local spin density vector. Inserting Eq.\,(\ref{SM_coupled_vectorial}) into Eq.\,(\ref{SM_cont}), we obtain the following set of differential equations for the charge and spin chemical potentials
\begin{subequations}
\label{decomposed_cont_fm}
\begin{align}
\label{sym_cont_fm}
&\frac{d^2}{dz^2}\mu_{c}(z) + p_{\sigma}\frac{d^2}{dz^2}\mu_{s}(z)=0\,,
\\
\label{antisym_cont_fm}
&\frac{d^2}{dz^2}\mu_{s}(z) - \frac{\mu_{s}(z)}{\lambda^2}=0\,,
\\
\label{antisym_cont_perp1}
&\frac{d^2}{dz^2}\mu_{s}^{\perp_1}(z) - \frac{\mu_{s}^{\perp_1}(z)}{\lambda_{\perp}^2}=0\,,
\\
\label{antisym_cont_perp2}
&\frac{d^2}{dz^2}\mu_{s}^{\perp_2}(z) - \frac{\mu_{s}^{\perp_2}(z)}{\lambda_{\perp}^2}=0\,,
\end{align}
\end{subequations}
where $\lambda=\sqrt{\sigma_{0}(1-p_{\sigma}^2)\tau_{sf}/2N(\epsilon_F)(1-p_N^2)}$ and $\lambda_{\perp}=\sqrt{\sigma_{0}\tau_{sf}/2N(\epsilon_F)(1-p_N^2)}$ are the spin diffusion lengths parallel and perpendicular to the magnetization, respectively. We thus find that the transverse spin chemical potentials are decoupled from $\mu_c$ and $\mu_s$ to leading order in the Hall angles, as are their boundary conditions \cite{kim2020generalized}. Therefore, for the purpose of UMR analysis,  we may disregard transverse spin chemical potentials.

\section{Contribution of Boundary Resistance}
\label{appendix_b}

The presence of a boundary resistance at the bilayer interface modifies the boundary conditions on the charge and spin chemical potentials as \cite{valet1993prb}
\begin{subequations}
\label{new_densities}
\begin{align}
\label{n(z)}
\mu_{c,F}\left(0^+\right) - \mu_{c,N}\left(0^-\right)
&=
- \gamma r_b \mathcal{J}_{z}\left(0\right),
\\
\mu_{s,F}\left(0^+\right) - \mu_{s,N}\left(0^-\right)
&=
r_b \mathcal{J}_{z}\left(0\right),
\end{align}
\end{subequations}
where $\gamma$ is the interfacial spin asymmetry coefficient and $r_b$ is the boundary resistance for a unit surface of the interface.  

Resolving the continuity equations presented in the main text with the modified boundary conditions, the total UMR coefficient $\zeta$ of the system may be calculated. This is comprised of the SH- and spin AH-UMR coefficients,  $\zeta = \zeta_{\text{SAH}} + \zeta_{\text{SH}}$. Taking the spatial average defined as $\bar{\zeta}\equiv \int_{-d_N}^{d_F} dz \;\zeta(z)/(d_N+d_F)$, to first order in the SH and spin AH angles, we obtain
\begin{widetext}
\begin{equation}
\label{UMR}
\bar{\zeta}
=
\frac{3 \left(p_{\sigma} - p_N\right)}{\epsilon_F} \left(\frac{\sigma _{0,F} \lambda_F}{\sigma _{0,F}d_F + \sigma _{0,N} d_N}\right)
\frac{\tanh \left(\frac{d_F}{\lambda_F}\right)\left[\theta_{\text{SAH}} \lambda_F \tanh \left(\frac{d_F}{2\lambda_F}\right) + \theta_{\text{SH}} \lambda_N \tanh\left(\frac{d_N}{2\lambda_N}\right)\right]}
{1+ \left(1-p_{\sigma}^2\right) \tanh \left(\frac{d_F}{\lambda_F}\right) \left[\left(\frac{\sigma_{0,F} \lambda_N}{\sigma_{0,N} \lambda_F}\right) \coth \left(\frac{d_N}{\lambda_N}\right) + \frac{r_b \sigma _{0,F}}{\lambda_F}\right]}\,,
\end{equation}
\end{widetext}
where, as in the main text, we note the relation between the spin AH- and SH-UMR coefficients
\begin{equation}
\frac{\bar{\zeta}_{\text{SAH}}}{\bar{\zeta}_{\text{SH}}}
=
\frac{\theta_{\text{SAH}} \lambda_F \tanh\left(\frac{d_F}{2\lambda_F}\right)}{\theta_{\text{SH}} \lambda_N \tanh \left(\frac{d_N}{2\lambda_N}\right)}\,.
\end{equation}
Taking into account the interfacial resistance, we have verified that, for a typical boundary resistance of $r_b \sim 1 \text{f}\Omega\;\text{m}$ \cite{valet1993prb,Bass07JP-CM_spin-diffusion}, the effect on the transport coefficients is negligible. Thus, we may safely neglect the interfacial resistance in the present study.

\bibliographystyle{nat_commune}
\bibliography{sahumr}
\end{document}